\documentclass[11pt,a4paper,final]{article}
 
\usepackage{graphicx}
\usepackage[breaklinks=true,colorlinks=true,linkcolor=blue,urlcolor=blue,citecolor=blue]{hyperref}
\usepackage{authblk}
\usepackage{sidecap}
\usepackage[cm]{fullpage}

	\sidecaptionvpos{figure}{c}
	
  \newcommand{\Omprime}{\ensuremath{\Omega^\prime{}}}
	\newcommand{\qprime}{\ensuremath{q^\prime{}}}
	\newcommand{\shat}{\ensuremath{\hat{s}}}

\begin{document}

\title{Using the local gyrokinetic code, GS2, to investigate global ITG modes in tokamaks. (I) $s-\alpha$ model with profile and flow shear effects}

\author[1]{P. A. Abdoul\thanks{paaa500@york.ac.uk}}
\author[1]{D. Dickinson}
\author[2]{C. M. Roach}
\author[1]{H. R. Wilson}
\affil[1]{York Plasma Institute, Department of Physics, University of York, Heslington, York, YO10 5DD, UK}
\affil[2]{CCFE, Culham Science Centre, Abingdon, Oxfordshire, OX14 3DB, UK}
\renewcommand\Authands{ and }
\maketitle

\begin{abstract}
\label{abstract}
This paper combines results from a local gyrokinetic code with analytical theory to reconstruct the global eigenmode structure of the linearly unstable ion-temperature-gradient (ITG) mode with adiabatic electrons. The simulations presented here employ the $s$-$\alpha$ tokamak equilibrium model. Local gyrokinetic calculations, using GS2 have been performed over a range of radial surfaces, $x$, and for ballooning phase angle, $p$, in the range $-\pi \le p \le \pi$, to map out the complex local mode frequency, $\Omega_{0}(x,p)=\omega_{0}(x,p)+i\gamma_{0}(x,p)$. Assuming a quadratic radial profile for the drive, namely $\eta_{i} = L_n/L_T$, (holding constant all other equilibrium profiles such as safety factor, magnetic shear etc.), $\Omega_{0}(x,p)$ has a stationary point. The reconstructed global mode then sits on the outboard mid plane of the tokamak plasma, and is known as a conventional or isolated mode, with global growth rate, $\gamma \sim$ Max$[\gamma_{0}(x,p)]$, where $\gamma_{0}(x,p)$ is the local growth rate. Taking the radial variation in other equilibrium profiles (e.g safety factor $q(x)$) into account, removes the stationary point in $\Omega_{0}(x,p)$ and results in a mode that peaks slightly away from the outboard mid plane with a reduced global growth rate. Finally, the influence of flow shear has also been investigated through a Doppler shift, $\omega_{0} \rightarrow \omega_{0} + n\Omprime x$, where $n$ is the toroidal mode number and $\Omprime$ incorporates the effect of flow shear. The equilibrium profile variation introduces an asymmetry to the growth rate spectrum with respect to the sign of $\Omprime$, consistent with recent global gyrokinetic calculations.
\end{abstract}

\section{Introduction}
\label{introduction}

Tokamaks~\cite{Wesson_1} provide one of the most stable and promising configurations for magnetic confinement fusion. However, confinement in tokamaks is not perfect. There are a number of mechanisms by which energy and particles can be transported across the magnetic flux surfaces from the core confinement region to the plasma edge. The main contribution is typically due to turbulent transport, which is widely believed to originate primarily from microinstabilities, driven by density and temperature gradients. The drift-type modes with low frequency compared to the cyclotron frequency, are the dominant tokamak microinstabilities~\cite{Wesson_1, Horton_1}, and the turbulence they derive often determines the minimum size of magnetic confinement fusion reactors, such us ITER and DEMO. Therefore, it is important to understand them and find a way to suppress them and/or reduce their effects. Previous theoretical studies have shown that flow shear can control the stability of drift waves, providing a mechanism to suppress or even stabilise them completely \cite{Dickinson_1, Connor_0, Terry_1, Kishimoto_1, Roach_1, Biglari_1, Waltz_1}. Furthermore, understanding the structure of these instabilities and associated threshold gradients is likely to be very important for developing simplified models for heat and particle transport. Microinstabilities are typically investigated via numerical simulations of the gyrokinetic equations~\cite{Rutherford_1, Frieman_1}~using both local and global approaches, as we now discuss. 

For high toroidal mode numbers, $n$, the rational flux surfaces (i.e. those where the safety factor, q, is rational) are closely packed so equilibrium quantities vary only weakly from one to the next. Two length scales can be identified: the equilibrium length scale, characterised by the minor radius, $r$, and  the distance between rational surfaces, $\Delta=1/n\qprime{}$ where a prime denotes derivative with respect to $r$. The separation between these length scales is exploited in ballooning theory~\cite{Taylor_1, Dewar_0} to perform an expansion in the small parameter, $\Delta/r$. All rational surfaces spanned by the mode are then equivalent to leading order, and local gyrokinetic codes, like GS2~\cite{Rewoldt_1, Dorland_1}, exploit this "ballooning symmetry to reduce the intrinsic 2D eigenvalue problem to 1D in the extended ballooning coordinate that is aligned with magnetic field lines, denoted by $\eta$. Local gyrokinetic codes do not directly yield the radial mode structure, but they do provide the structure along the magnetic field line, together with the local mode eigenvalues, $\Omega_0(x,p)$, depending on radial coordinate $x=\frac{r-r_{0}}{a}$, where $a$ is the minor radius of the tokamak plasma, and the ballooning angle parameter, $p$. At this order, both $x$ and $p$ are free parameters in the local gyrokinetic simulations. However, they are constrained in the higher order ballooning theory, which then yields the radial mode structure and global eigenvalue, $\Omega$ 
(which is a complex mode frequency).

In this paper, we shall reconstruct the global mode structures purely from local solutions, employing the Fourier-Ballooning representation~\cite{Mahajan_1} described in section~\ref{technique}. According to this equation, the global mode structure is related to both the local mode structure $\xi_{x}(p, \theta, t)$ and its corresponding complex mode frequency $\Omega_{0}(x,p)$. It has previously been established that under the very special conditions where $\Omega_{0}(x,p)$ has a stationary point, a so-called isolated or conventional ballooning mode is obtained from the higher order analytical theories. This type of mode, originally studied in the context of ideal MHD theory~\cite{Taylor_2}, is usually located at the outboard mid plane, where the poloidal angle $\theta=0$~\cite{Dickinson_1}. Profile variations, however,  are usually present in realistic situations, and global gyrokinetic simulations do not always find isolated modes. For example, global gyrokinetic simulations of the linear electrostatic ion temperature gradient (ITG) mode in ASDEX Upgrade plasmas, predicted that the mode was slightly shifted relative to the outboard mid plane~\cite{Bottino_1}. However, in exceptional situations, e.g. when a critical value of flow shear compensates the effect of profile variation, an isolated mode can exist~\cite{P.Hill_1}. More generally, linear global gyrokinetic simulations have shown that the radial profile variations break the up-down poloidal symmetry of the mode, providing a mechanism for toroidal momentum transport~\cite{Camenen_1}. This transport influences so-called intrinsic rotation~\cite{Peeters_0}, which is of crucial importance in generating plasma flows in a machine like ITER for which external torque is very small. 

Using a simplified fluid model of ITG modes, it has been shown that global mode structures can be accurately constructed from local ballooning results for given radial profiles~\cite{Dickinson_1}. Our work builds on that by applying the theory to a more realistic and generally applicable kinetic plasma model. Specifically the global mode structures of an ion temperature gradient driven microinstability have been reconstructed directly from local gyrokinetic solutions from the GS2 code. To achieve this, numerical solutions from GS2, have been used with the higher order analytical theory presented in reference~\cite{Taylor_1}. We have carried out these calculations using the $s-\alpha$ equilibrium model~\cite{Connor_1, Lapillonne_1}, in which circular magnetic flux surfaces and high aspect ratio have been assumed.

This paper is organised as follows. Section~\ref{technique} presents the theoretical formalism on which this paper is based. The results, from applying this technique to our example problem, are presented in section~\ref{results}. Finally, section~\ref{conclusion}~summarises our conclusions, and plans for future work.
\section{The technique: From local to global gyrokinetic calculations}
	\label{technique}

We employ the initial value local gyrokinetic code, GS2~\cite{Rewoldt_1, Dorland_1}, to solve the linearised gyrokinetic equations numerically~\cite{Rutherford_1, Frieman_1}. GS2 provides the mode structure along a given magnetic field line with its local eigenvalue, $\Omega_{0}(x,p)$ for a prescribed radial coordinate, $x=\frac{r-r_{0}}{a}$, and ballooning angle, $p$. Here, $r$ is the minor radius, and $x$ is a normalised distance from the rational surface at $r=r_{0}$, with $r=a$ being the plasma edge. To investigate the linear global mode properties by using only solutions from the local gyrokinetic code, we employ the Fourier-Ballooning (FB) representation~\cite{Mahajan_1}
	
	\begin{equation}
\label{equ_1}
\phi(x,\theta, t) = \int_{-\infty}^{\infty}\xi_{x}(p, \theta, t)\exp(-inq_{0}\theta)\exp(-in\qprime{}x(\theta - p))A(p) dp
\end{equation}

Here, $\phi(x,\theta, t)$ is the global mode. The parameter $q_{0}$ is the safety factor evaluated at $x=0$. The function $\xi_{x}(p,\theta,t)$, obtained from the local gyrokinetic code, GS2, represents the local mode structure where the poloidal angle, $\theta$ measures the distance along a field line. Note that $\xi$ is invariant under the transformation $p\rightarrow p+2\pi l$ and $\theta \rightarrow \theta+2\pi l$ for any integer $l$. The envelope $A(p)$ is required to reconstruct the radial mode structure and is obtained from the higher order theory. From equation(\ref{equ_1}) and the aforementioned symmetry properties of $\xi$, $A$ must be periodic in $p$ in order for $\phi$ to be periodic in $\theta$.

We seek a global eigenmode satisfying:
	\begin{equation}
\label{equ_1a}
\frac{d\phi(x,\theta, t)}{dt} = -i \Omega \phi(x,\theta, t)
\end{equation}
Where, $\Omega=\omega+i\gamma$ represents the true, global complex mode frequency, with $\omega$ and $\gamma$, corresponding to the frequency and growth rate, respectively. Substituting equation(\ref{equ_1}) into (\ref{equ_1a}), and noting $\frac{d\xi_{x}(p, \theta, t)}{dt} = -i \Omega_0(x,p) \xi_{x}(p, \theta, t)$ for the local eigenfunction $\xi$ with local complex mode frequency $\Omega_{0}(x, p)$, gives:
\begin{equation}
 \label{equ_1b}
\Omega \phi(x,\theta, t) = \int_{-\infty}^{\infty} \Omega_0(x,p)\xi_{x}(p, \theta, t)\exp(-inq_{0}\theta)\exp(-in\qprime{}x(\theta - p))A(p) dp
\end{equation}
Note that, the local complex mode frequency $\Omega_{0}(x, p)=\omega_{0}(x, p) + i\gamma_{0}(x, p)$, where $\omega_{0}$ and $\gamma_{0}$ are the local frequency and growth rate, respectively, is measured in unit of $(v_{ti}/R)$. It follows that time is measured in unit of $(R/v_{ti})$. Using equation(\ref{equ_1}) for $\phi$, we can then write our eigenmode condition in the form:
\begin{eqnarray}
 \label{equ_1c}
\int_{-\infty}^\infty (\Omega - \Omega_{0}(x,p)) \xi(p,\theta,t) \exp(-inq_{0}\theta) \exp(-inq^\prime{}x(\theta - p)) A(p)dp=0
\end{eqnarray}
Here, the local complex mode frequency, $\Omega_{0}(x, p)$, is mapped out by running the local gyrokinetic code, GS2, many times over the required range of $x$ and a period of $p$. We then Taylor expand in $x$ and Fourier expand in $p$ to derive:
\begin{equation}
\label{equ_2}
\Omega_{0}(x,p)=\sum_{k=0}^{N_{k}}\sum_{m=0}^{2}a_{k}^{(m)}x^{m}cos(kp)
\end{equation}
where $N_{k}$ is the number of Fourier modes retained. The coefficients $a_{k}^{(m)}$ are complex numbers, derived by fitting to the $\Omega_{0}(x,p)$ results from GS2. In the limit of high toroidal mode number, $n \rightarrow \infty$, exploited by local gyrokinetic codes, the radial extents of the reconstructed global modes are very small compared to the equilibrium scale lengths. Therefore, we do not expect the complex mode frequency to have radial components higher than second order, i.e $m \leq 2$. Now substituting our fit for $\Omega_0 (x, p)$ from equation(\ref{equ_2}) into equation(\ref{equ_1c}), we have
\begin{eqnarray}
\label{equ_2a}
\sum_{k=0}^{N_k} \int_{-\infty}^\infty & \left[ (\Omega - a_{k}^{(0)} \cos(kp))A - \frac{i}{n\qprime{}} a_{k}^{(1)} \cos(kp)\frac{dA}{dp}+\frac{1}{(n\qprime{})^2} a_{k}^{(2)} \cos(kp) \frac{d^{2} A}{dp^{2}} \right] \nonumber \\
& \xi_{x}(p,\theta,t) \exp(-inq_{0}\theta)\exp[-in\qprime{}x(\theta - p) ] dp = 0
\end{eqnarray}
where we have integrated by parts and assumed that $A(p)$ has the fastest variation with $p$ to derive the mapping $x^{k} A \rightarrow (-i/n\qprime{})^{k} d^{k}A/dp^{k}$. Equation(\ref{equ_2a}) must hold for all $\theta$, which then provides our final equation for $A(p)$:
\begin{equation}
\label{equ_4}
 a_{k}^{(2)}\cos(kp) \frac{1}{(n\qprime{})^2}\frac{d^{2} A}{dp^{2}} - a_{k}^{(1)}\cos(kp)\frac{i}{n\qprime{}}\frac{dA}{dp}+[\Omega - a_{k}^{(0)}\cos(kp)]A = 0
\end{equation}
This is to be solved subject to a periodicity boundary condition to determine the eigenfunction $A(p)$ and the global complex mode frequency, $\Omega$ as an eigenvalue. Knowledge of $A(p)$, together with $\xi_{x}(p , \theta , t)$ from GS2 then allows the full 2D eigenfunction, $\phi(x, \theta, t)$, to be reconstructed from equation(\ref{equ_1}). 

Equation(\ref{equ_4}) can be solved analytically in two limiting cases: either $a_{k}^{(1)}=0$ or $a_{k}^{(2)}=0$ for all $k$; the former implies that $\Omega_{0}(x, p)$ is stationary at $x=0$, i.e. $\partial \Omega_0 /\partial x |_{x=0} = 0$. These limits were considered in references~\cite{Dickinson_1, Taylor_1}, for example, to derive the two different eigenmode classes which they referred to as `isolated modes' in the special cases when $a_{k}^{(1)} = 0$ and 'general modes' in the more usual situation that $a_{k}^{(1)} \neq 0$ (in which case the terms in $a_k^{(2)}$ can often be neglected as they are small in the limit of large $n$). In this paper, we shall retain both terms and solve equation(\ref{equ_4}) numerically.


\section{Results}
	\label{results}
\begin{figure}[t!]
		\centering
\includegraphics[width=1.0\textwidth]{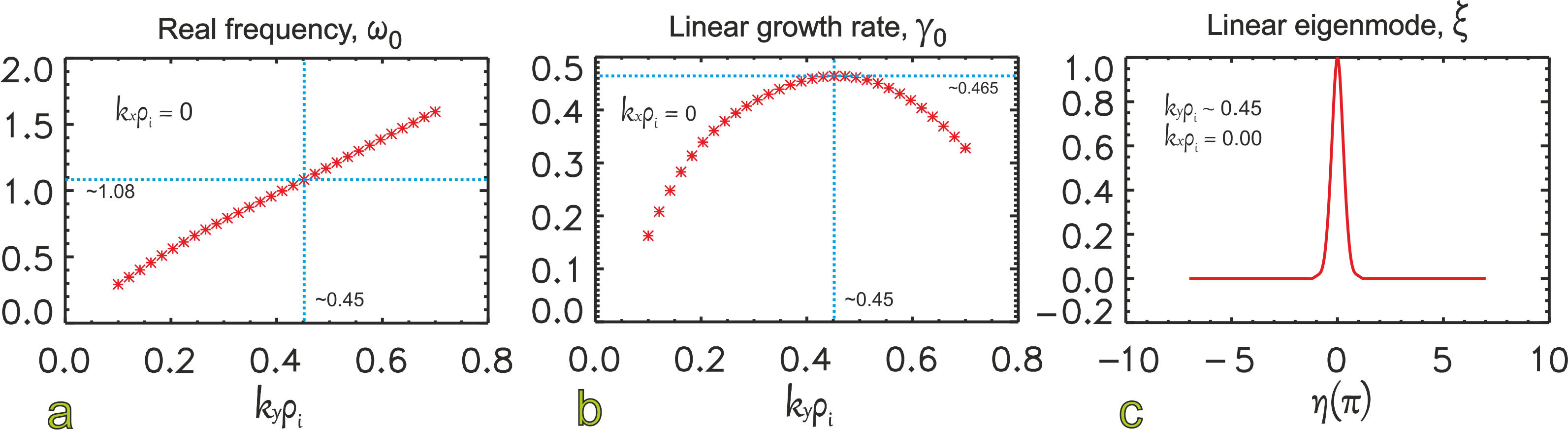}
\caption{\label{fig_1}From left to right: The variation of (a) real frequency, $\omega_{0}$, (b) linear growth rate, $\gamma_{0}$, with $k_{y} \rho_{i}$ and (c) the local mode structure, $\xi(\eta, p=0)$, for the most unstable mode ($k_{y} \rho_{i}=0.45$), along the magnetic field line, $\eta$. Note that, both $\omega_{0}$ and $\gamma_{0}$ are measured in unit of $(v_{ti}/R)$ and these simulations have been carried out at mid-radius, i.e $x=0$ ($r=r_{0}$).}
\end{figure}
 \begin{SCfigure}
		\centering
 \includegraphics[width=0.50\textwidth]{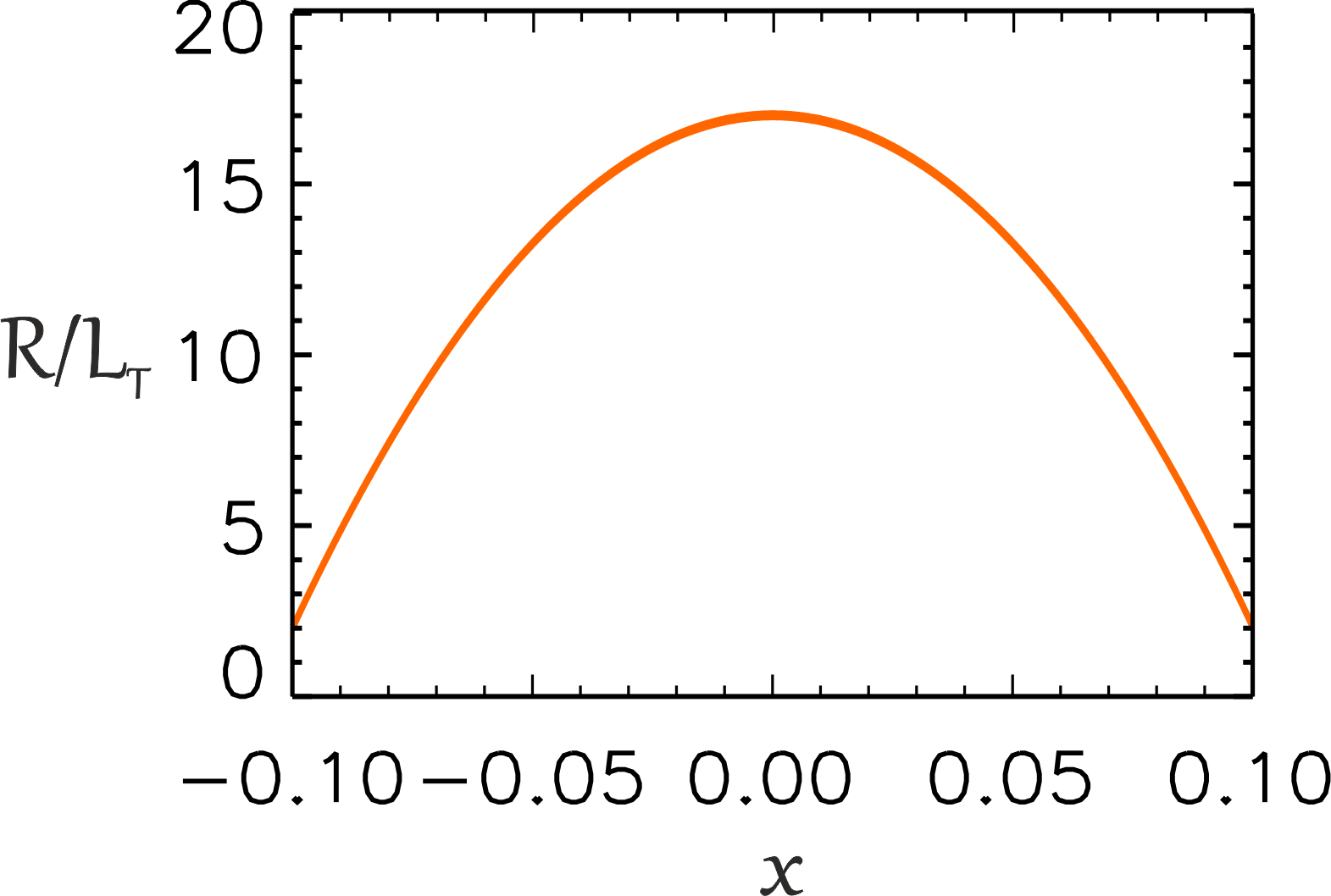}
 \caption{\label{fig_2}Radial profile for the temperature gradient, $R/L_{T} = 17.00 - 1500 x^{2}$. Note that, $x$, normalized to minor radius $a$, measures the distance from a particular mode rational surface located at the centre of the domain.}
\end{SCfigure}
Before considering the global calculations, we first start with the local ballooning analysis. Here, only Ion Temperature Gradient (ITG) modes~\cite{Wesson_1, Horton_1, Romanelli_1, Peeters_1}, believed to be one of the causes for turbulent transport in tokamak plasmas, have been considered. In this paper for convenience we have limited ourselves to linear electrostatic calculations with adiabatic electrons. The model equilibrium parameters that we have used are given in table~\ref{table_1}. Fig~\ref{fig_1} shows the real frequency, $\omega_{0}$, and growth rate, $\gamma_{0}$, as a function of $k_{y}\rho_{i}$ for the mode with $p=0$, together with the local mode structure, $\xi$, in ballooning space, $\eta$. To map from the infinite domain, $\eta$, to the poloidal angle, $\theta$, we make use of the relation $\xi(\theta + 2l\pi, p + 2l\pi)=\xi(\theta, p)$ for any integer $l$. Here, $k_{y}$ is the binormal perpendicular wave number and $\rho_{i}$ is the ion Larmor radius. 
The most unstable mode is found at $k_{y}\rho_{i}=0.45$. In the following, we apply the technique presented in section~$\ref{technique}$~to reconstruct the global mode structures for this most unstable mode.
\begin{table}[t] 
\caption{The model parameters that have been used throughout this paper. Note that subscript 0 indicates that the parameter has been evaluated at the rational surface $r=r_{0}$.} 
\centering 
\begin{tabular}{||c||c|c|c|c|c|c|c|c|c|c|c|c||c||} 
\hline\hline 
Parameter & $\shat$ & $q_{0}$	& $R/L_{T}$ & $R/L_{n}$ & $k_{y}\rho_{i}$ & $\epsilon$ & a & R & $\beta$ & $n\qprime$ & $\nu_{ii} \times \frac{R}{v_{ti}}$ & $\frac{T_{i}}{T_{e}}$ \\ [0.5ex] 
\hline\hline 
Value & 1.5 & 1.4 & 17.0 & 2.2 & 0.45 & 0.18 & 0.27m & 1.50m & 0.0 & 300 & 0.8 & 1.0 \\ [0.2ex] 
\hline\hline 
\end{tabular} 
\label{table_1} 
\end{table} 

\subsection{Global Calculations: The quadratic \texorpdfstring{$\eta_{i}$}{Lg} profile}
	\label{global1}
\begin{figure}[t!]
	\centering
	\includegraphics[width=1.0\textwidth]{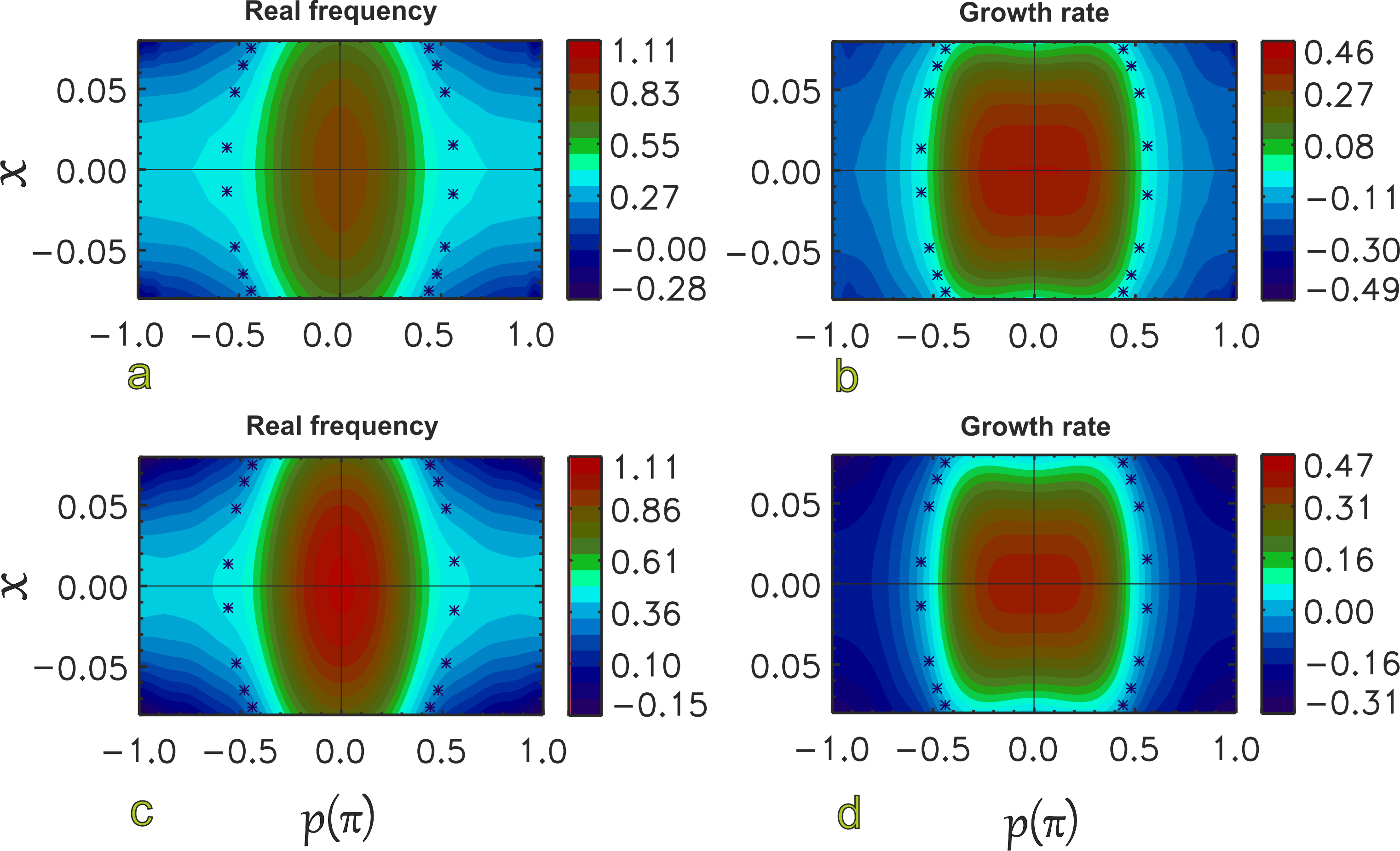}
\caption{\label{fig_3}Contour plots of real and imaginary parts of the local complex mode frequency, measured in unit of $(v_{ti}/R)$, as a functions of radius, $x$, and ballooning angle, $p$, for quadratic $R/L_{T}$ radial profile, while excluding other profile variations. (a) and (b) are, respectively, the frequency and growth rate obtained from the local gyrokinetic code, GS2. The corresponding frequency and growth rate from the fitted model, using $\Omega_{0}(x,p)=\sum_{k=0}^{10} \sum_{m=0}^{2} a_k^{m} cos(kp)$, are presented in (c) and (d) respectively. The $\ast$ symbol indicate the marginal stability contour.} 
\end{figure}	
\begin{figure}[t!]
	\centering
	\includegraphics[width=1.0\textwidth]{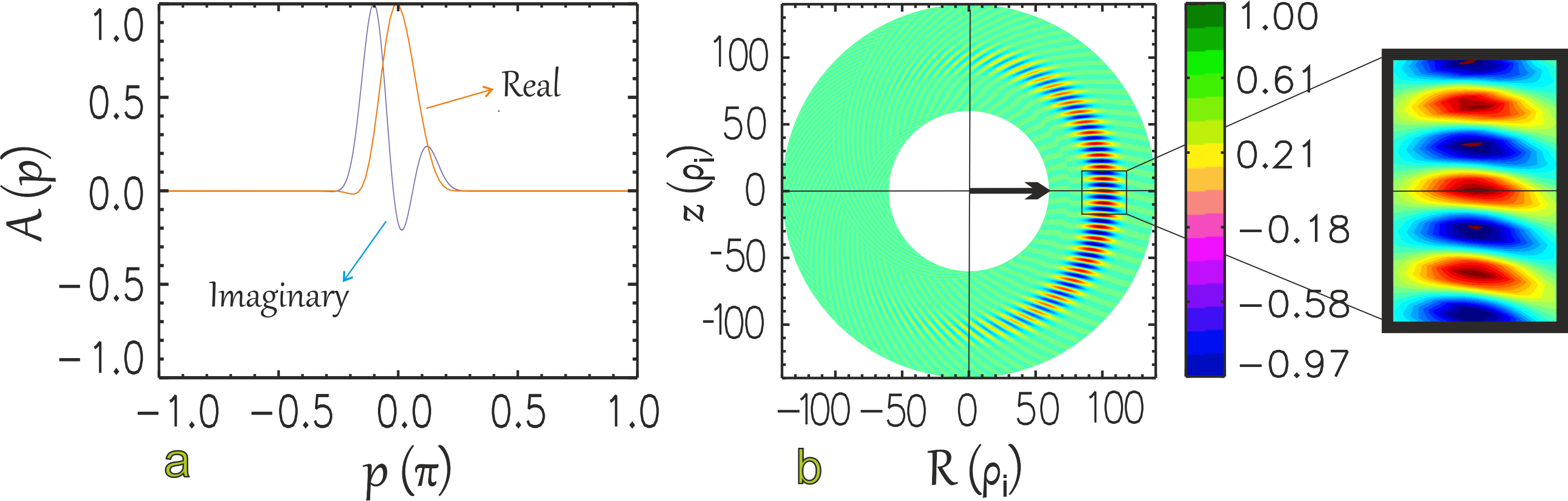}
\caption{\label{fig_4}From left to right: (a) presents the numerical solution of the envelope function, $A(p)$, obtained from equation($\ref{equ_4}$), as a function of ballooning angle, $p$, and (b) reconstructed electrostatic potential global mode structure, $\phi(x, \theta)$, in the poloidal plane. Note that profile variations other than $L_{T}$ are excluded, and the mode structure, centred on the outboard mid plane, is aligned radially where $\theta = 0$.}
\end{figure}
\begin{SCfigure}
	\centering
	\includegraphics[width=0.50\textwidth]{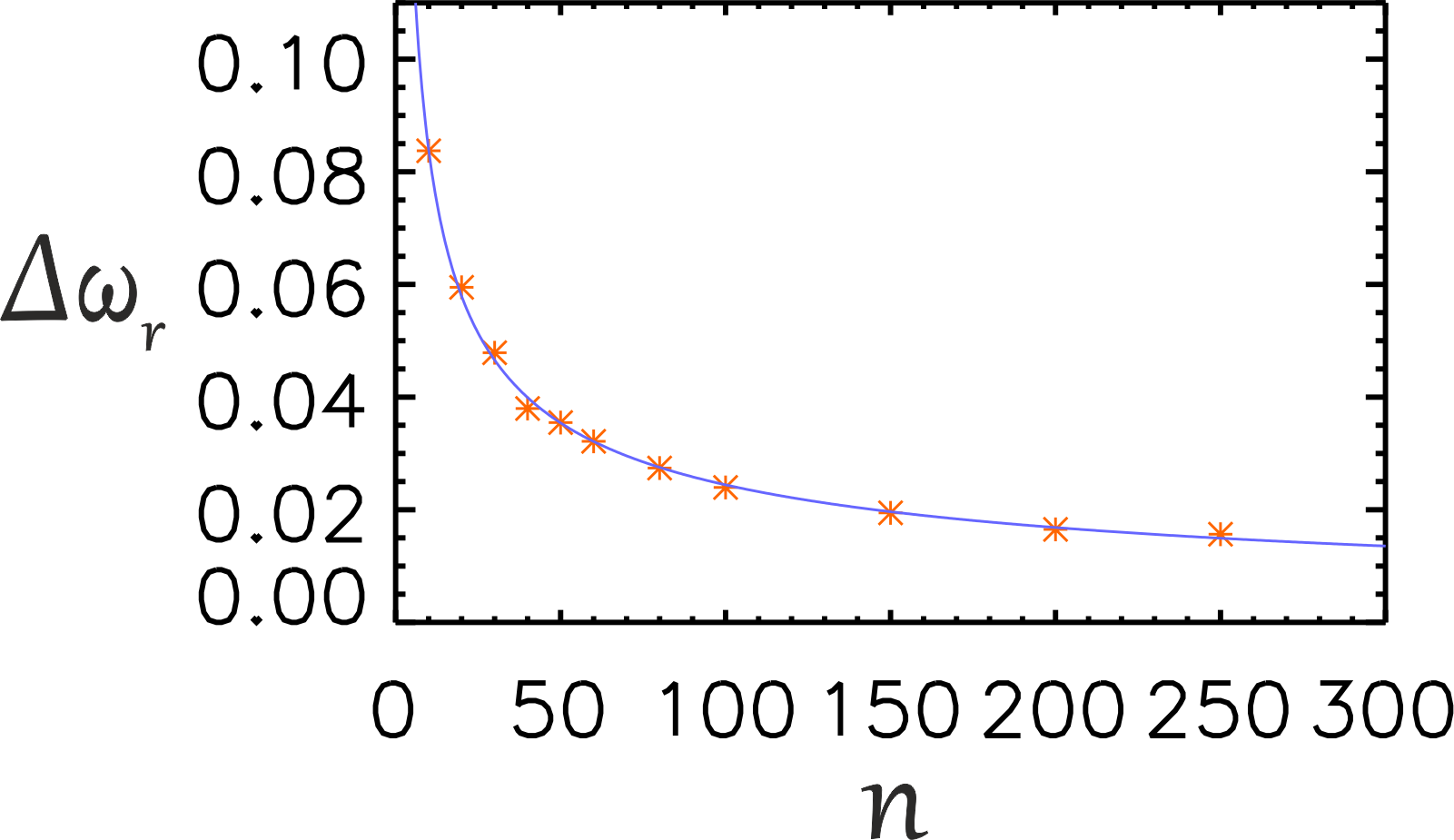}
	\caption{\label{fig_5}The radial mode width, $\Delta w_{r}$, as a function of toroidal mode number, $n$. It scales, approximately,  inversely with square root of $n$ according to: $\Delta w_{r} = 0.289n^{-0.54}$.}
\end{SCfigure}
	\begin{figure}[t!]
	\centering
	\includegraphics[width=1.0\textwidth]{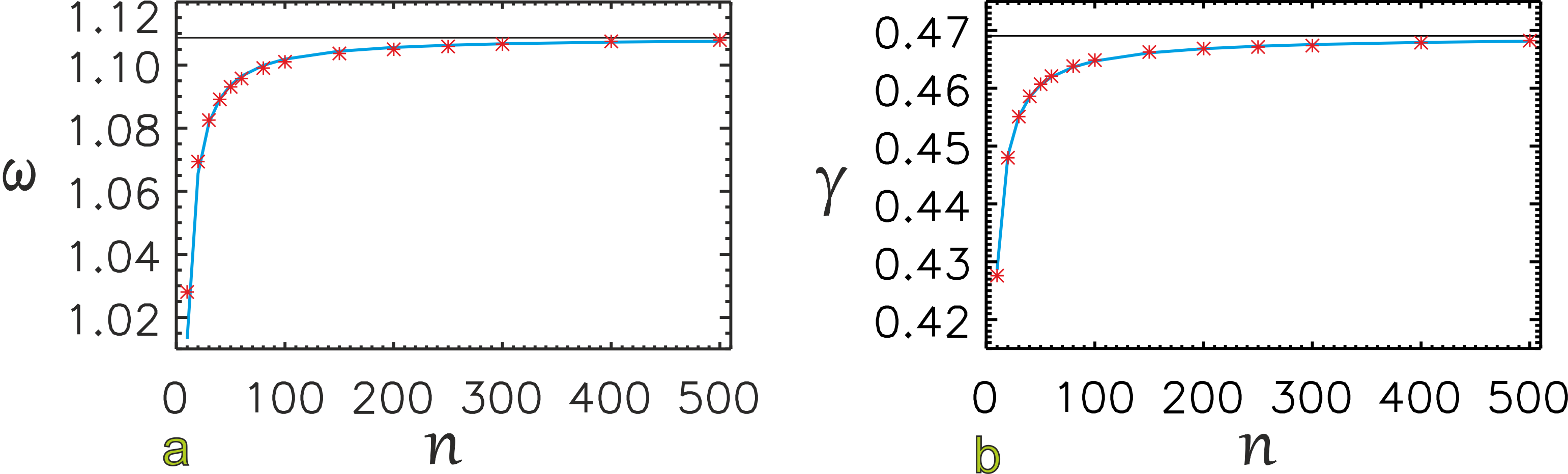}
\caption{\label{fig_6}The global real frequency $\omega$ (a) and growth rate $\gamma$ (b) as functions of toroidal mode number, $n$. The $\omega$ and $\gamma$ are measured in unit of $(v_{ti}/R)$. They scale with $n$ as follow: 
$\omega = \omega_{0}(0,0) - 1.35 n^{-1.15}$, $\gamma = \gamma_{0}(0,0) - 377100 n^{-0.97}$, where $\Omega_{0}(0,0)=\omega_{0}(0,0)+i\gamma_{0}(0,0) = 1.1086+i0.4690$ is the value of local complex mode frequency, indicated by two horizontal lines, calculated at radial position $x=0$ and ballooning angle $p=0$. Note that, as $n$ increases the value of $\Omega$ converges to $\Omega_{0}$.}
\end{figure}

To reconstruct the global mode structure, we first consider a quadratic radial temperature gradient profile, $R/L_{T} = 17.00 - 1500 x^{2}$, that peaks radially at the centre of the domain, i.e at $x=0$, see Fig $\ref{fig_2}$. We hold all other equilibrium profiles constant at those values given in table~\ref{table_1}. In particular we have assumed that $R/L_{n}=2.2$ over the entire simulation so that the profile variation in the driving source, $\eta_{i}$, is determined by the $R/L_{T}$ profile. GS2 calculations across the radial domain, $x$, and full range in ballooning angle $-\pi \leq p \leq \pi$, then provide the local complex mode frequency, $\Omega_{0}(x, p)$. Fig $\ref{fig_3}$ shows the contour plot for the $\Omega_{0}(x,p)$ numerical data, taken from GS2, and the model fit using the parametrization given in equation(\ref{equ_2}). As expected, due to the quadratic radial variation of the temperature gradient profile $R/L_{T}$, the coefficients with $m=1$ are all zero. To obtain $A(p)$ and its eigenvalue $\Omega$, we solve equation(\ref{equ_4}), numerically, with a periodic boundary condition, i.e $A(p-\pi)=A(p+\pi)$. Note that including Fourier harmonics above $N_{k} \geq 7$, the global eigenvalue converges to a constant value, $\Omega = \omega+i\gamma = 1.093 +0.4609i$. For this reason, in what follows, we have used $N_{k}=10$. 
	
Using this numerical solution for $A(p)$, and the solution for $\xi(x,p,\theta)$ from GS2, equation($\ref{equ_1}$) is solved  to obtain the global mode structure, $\phi(x, \theta)$. The function $A(p)$ and the corresponding solution for $\phi(x,\theta)$ in the poloidal cross-section are shown in Fig~\ref{fig_4}. We see that $A(p)$ is highly localized about $p=0$. This leads to a mode that balloons on the outboard mid-plane, where $\theta=0$, with growth rate of $\gamma=$Max$[\gamma_{0}(x,p)] = 0.4609$. This mode is the isolated mode identified in~\cite{Dickinson_1, Taylor_1, Dewar_0}. Isolated modes are only possible when both the local real frequency, $\omega_{0}(x, p)$, and the local growth rate, $\gamma_{0}(x, p)$ are stationary at the same radial position, i.e where $x=0$ for our case, see Fig~\ref{fig_3}.

The radial mode width, $\Delta w_{r}$, and its variation with the toroidal mode number, $n$, has also been calculated and presented in Fig $\ref{fig_5}$. We have taken the magnitude of the radial slice from the constructed mode structure at $\theta=0$, where the mode peaks, i.e $\phi(x,0)$ and then fitted a Gaussian function to it. The radial width, $\Delta w_{r}$, has then been calculated as the full width at half maximum, FWHM, of the Gaussian. We find that the radial width of the mode scales inversely with the square root of $n$, $\Delta w_{r} \sim n^{-0.54}$. This is expected for these isolated modes~\cite{Taylor_2}.

Changing the toroidal mode number also affects the global mode frequency, $\Omega$.  Fig $\ref{fig_6}$ shows that both the real frequency, $\omega$, and the linear growth rate, $\gamma$, scale with $n$ according to: $\omega = \omega_{0}(0,0) - 1.35n^{-1.15}$ and $\gamma = \gamma_{0}(0,0) - 377100n^{-0.97}$, respectively. With $\omega_{0}(0,0) = 1.1086$ and $\gamma_{0}(0,0) = 0.4690$, respectively, being the values of the local frequency and growth rate calculated at $x=0$ and $p=0$. This indicates that the finite $n$ correction scales inversely with the toroidal mode number as expected from conventional ballooning theory~\cite{Taylor_2}, i.e $\sim n^{-1}$. In the limit of $n$ goes to $\infty$, the global growth rate is equal to the local growth rate calculated at $p=0$ and $x=0$, i.e $\gamma=$Max$[\gamma_{0}(x,p)]$. This is only true for this special case where $\Omega_{0}(x,p)$ has a stationary point at $x=0$. Our results are consistent with the predictions of analytical theory of the higher order ballooning calculations presented elsewhere in the literature, e.g in references~\cite{Taylor_1, Dickinson_1, Romanelli_1}. As we shall see in the following section, taking into account the radial variation of other equilibrium profiles, e.g safety factor $q$, introduces a small, but significant, deviation from conventional ballooning theory.

\subsection{Global Calculations: Profile variations and shear flow effects}
	\label{global2}
	
\begin{figure}[t!]
 \centering
\includegraphics[width=1.0\textwidth]{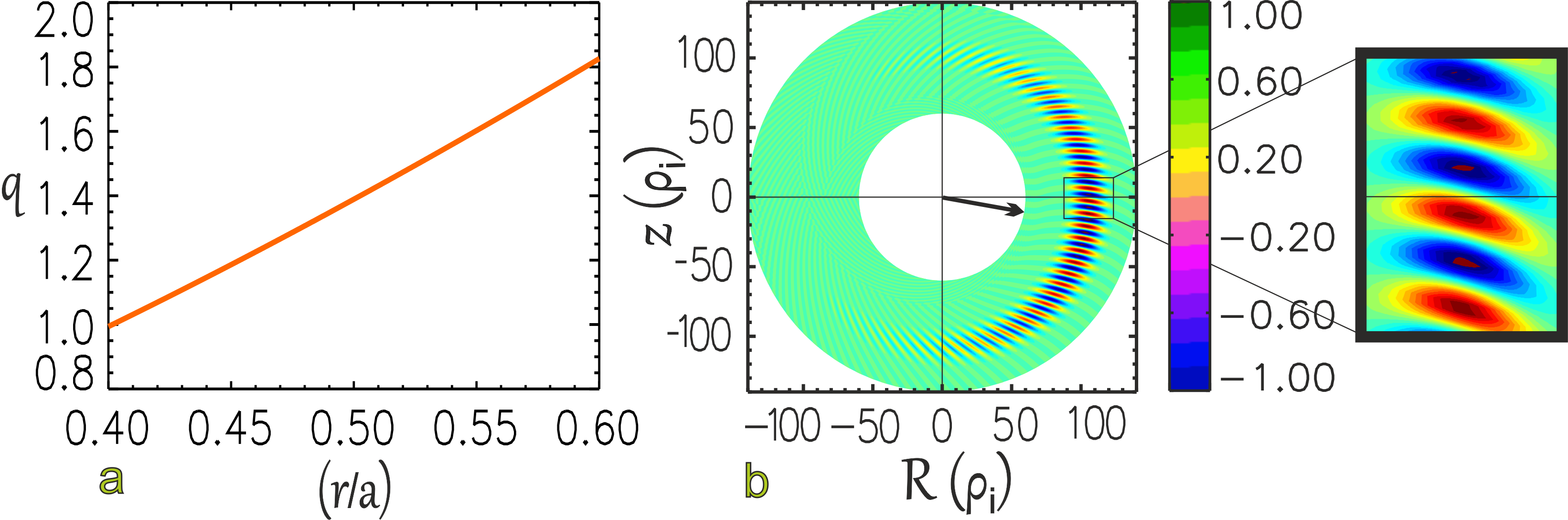}
\caption{\label{fig_7}From left to right: (a) Radially varying $q$ profile, $q=3.923(r/a)^{\shat}$, with constant magnetic shear $\shat=1.5$, and (b)  reconstructed global mode in the poloidal cross-section, $\phi(x,\theta)$. Note that the mode is now tilted on the outboard mid plane, where $\theta = 0$, due to the profile variation (c.f. Fig~\ref{fig_4}).}
 \end{figure}
	
We repeat the previous analysis, but now introduce a radially varying safety factor profile, $q=3.93(r/a)^{\shat}$, with constant magnetic shear parameter $\shat=1.5$. The $q$ profile varies over the length scale of the instability and gives rise to coefficients which are linear in $x$ into equation($\ref{equ_2}$). This modifies the local mode frequency, $\Omega_{0}(x,p)$, i.e. introduces terms with $m=1$ in the Taylor expansion of $\Omega_{0}(x,p)$. Furthermore, this profile variation affects the real frequency, $\omega_{0}(x,p)$, and the linear growth rate, $\gamma_{0}(x,p)$, differently, breaking their symmetry about $x=0$. It introduces a radial shift in the position of the maximum of $\omega_{0}(x,p)$ with respect to $\gamma_{0}(x,p)$, which affects the poloidal position of the mode peak and its stability.

Fig~\ref{fig_7} presents both the radially varying $q$ profile and the reconstructed global mode structure in the poloidal cross-section, $\phi(x,\theta)$. The mode shifts slightly downward with respect to the out-board mid plane, breaking the up-down poloidal symmetry with a slight reduction, $\approx 3\%$, in the growth rate, compared to the flat $q$ profile case of the previous section. As we can see, the radial $q$ profile variation tilts the structure on the outboard mid plane. This finding agrees qualitatively with other published results for full global simulations of linear ITG modes presented in~\cite{Camenen_1}. They have shown that the radial profile variation can induce toroidal momentum transport in the plasma, and this transport consequently can influence intrinsic rotation~\cite{Peeters_0} which is of particular interest for a machine like ITER for which the external source of torque will be negligible.    

\begin{figure}[t!]
	\centering
	\includegraphics[width=1.0\textwidth]{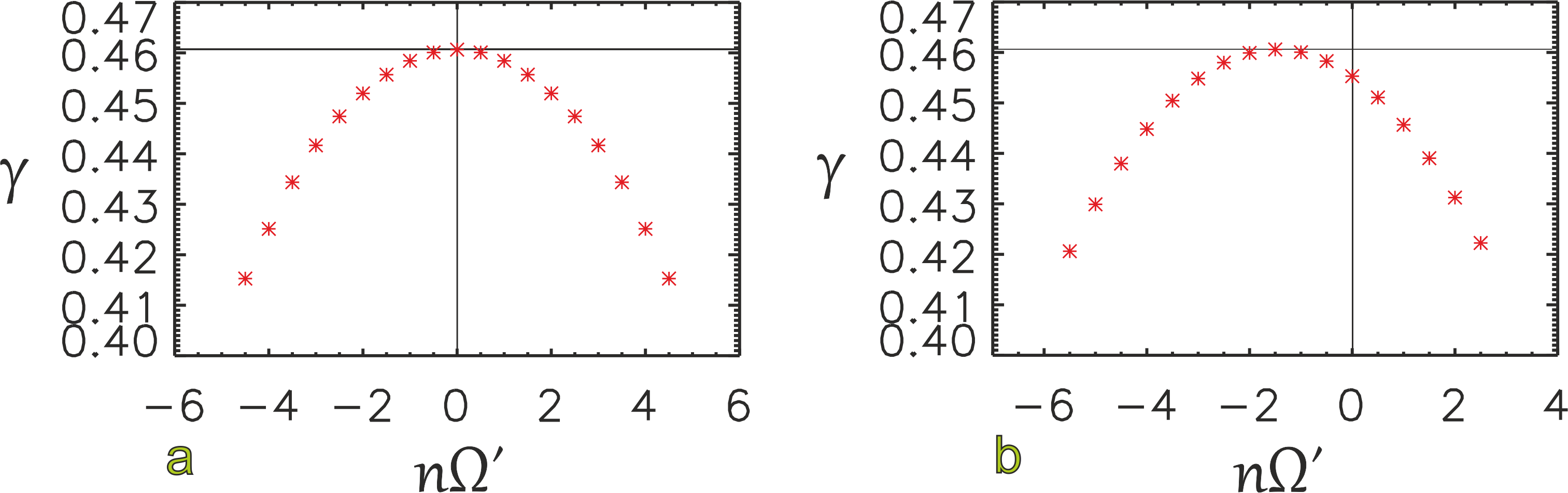}
	\caption{\label{fig_8}The linear growth rate, $\gamma$, is measured in unit of $(v_{ti}/R)$, as a function of flow shear, $n\Omprime$, calculated for the most unstable mode with $k_{y}\rho_{i}=0.45$. The toroidal mode number, $n=50$ and $\qprime \approx 6$. The radial variation of the $q$ profile is excluded in (a) and included in (b).}
\end{figure}
\begin{figure}[t!]
	\centering
	\includegraphics[width=1.0\textwidth]{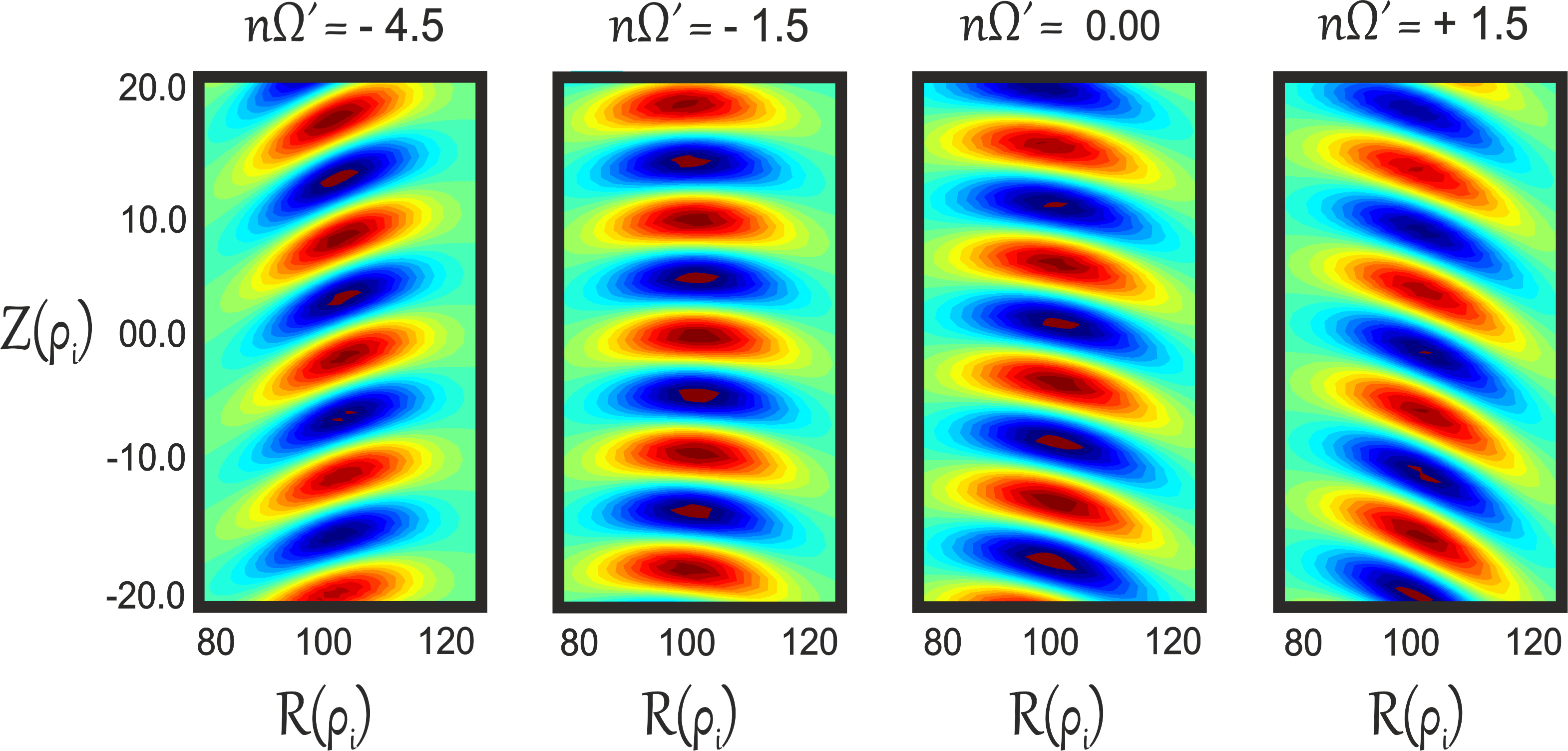}
	\caption{\label{fig_9}Reconstructed global mode structure in a small region of the poloidal cross-section at the outboard mid plane, under the combined effects of both radially varying $q$-profile and flow shear, $n\Omprime$. From left to right, $n\Omprime=-4.5,-1.5, 0$ and $+1.5$, respectively. For $n\Omprime=0$, the mode is already tilted, due to the profile variation effect, while a critical value of flow shear, occurs at $n\Omprime\cong-1.5$, cancels out the effect of profile variation and, once again, the mode structure is aligned radially (as for a conventional ballooning mode). From~\ref{fig_8}, we see that this is the maximally unstable flow shear.}
\end{figure}     

We have also studied the influence of a small, constant flow shear on the reconstructed mode structures and their stability, by introducing a Doppler shift in $\Omega_{0}$,
\begin{equation}
\label{equ_5}
\Omega_{0}(x,p) \rightarrow \Omega_{0}(x,p) +  n\Omprime x = (\omega_{0}(x,p) + n\Omprime x) +i\gamma_{0}(x,p),
\end{equation} 

where $\Omprime$ is constant and controls the flow shear effect. Thus $\Omprime x$ represents the rotation frequency of the magnetic flux surfaces measured relative to the mode rational surface at $x=0$ where the reconstructed global mode sits. We have considered flow shear cases, with $-4.0 \leq n\Omprime \leq +4.0$. 

From equation\ref{equ_5} it is clear that the flow shear introduces linear, $m=1$, terms into the $a_{k}^{(m)}$ coefficients of equation\ref{equ_2}. This indicates that both flow shear and the radial $q$-profile variation are expected to have similar effects on the global mode structures and their instabilities. 

Depending on the sign of the flow shear the reconstructed global mode shifts upward or downward relative to the outboard mid plane with a structure in the poloidal plane that is very similar to that in Fig~\ref{fig_7}. For a mode peaking at $\theta=0$ in the absence flow shear, increasing flow shear tilts the mode structure on the outboard mid plane and lowers the linear growth rate. Fig~\ref{fig_8} shows how the growth rate varies with flow shear. Excluding the $q$-profile variation, see Fig~\ref{fig_8}a, the growth rate curve is symmetric about $\Omprime=0$. However, the $q$-profile variation introduces an asymmetry to the growth rate with respect to the sign of flow shear, as shown in Fig~\ref{fig_8}b. The effect of this asymmetry on the reconstructed global mode structure is illustrated in Fig~\ref{fig_9}. For $n\Omprime=0$ the structure is already tilted, but increasing flow shear in the negative direction acts to re-align the mode radially. For a critical value of flow shear, $n\Omprime \cong -1.5$, the effect of the profile variation is completely compensated, allowing an isolated mode again to form with largest growth rate, $\gamma=$Max$[\gamma_{0}(x,p)]$. Increasing flow shear even further, below the critical value, tilts the mode structure in the negative direction and lowers its linear growth rate again. These results obtained purely from solutions of GS2, are in very good qualitative agreement with global calculations of linear electrostatic ITG modes presented in reference {\cite{P.Hill_1}}. 

For more realistic cases where we take other profile variations into account we do not in general expect an isolated mode to be observed in the global calculations, except for a critical value of flow shear. This is a special case, and generally the linear growth rate of microinstabilities will be lower.
\section{Conclusion}
	\label{conclusion}
	In this work we have reconstructed radial and poloidal mode structures for linearly unstable electrostatic ITG microinstabilities, purely from solutions of the local gyrokinetic code, GS2. Firstly, we introduce a quadratic radial profile for the mode driving source, while all other equilibrium parameters are held constant. We found that the complex mode frequency $\Omega_{0}(x,p)$ is stationary at the mode rational surface, at $x=0$, where the global mode sits. This leads to an isolated type of mode that peaks at the outboard mid plane with a large growth rate, $\gamma \sim$ Max$[\gamma_{0}(x,p)]$. These results are in very good qualitative agreement with the simplified fluid model of ITG modes presented in {\cite{Dickinson_1}}. Moreover, introducing a radial variation in the $q$ profile, we have found that the radial position of the stationary points in both local frequency, $\omega_{0}$ , and growth rate, $\gamma_{0}$, are slightly shifted with respect to each other. In this case, the reconstructed global mode becomes less unstable and shifts slightly downward with respect to the outboard mid plane. A constant flow shear, introduced as a Doppler shift in the real frequency, influences the global mode, reducing its linear growth rate and tilting the mode structure at the out board mid plane. Taking the $q$-profile variation effect into account causes an asymmetry in the growth rate with respect to the sign of flow shear. These findings are qualitatively similar to those captured in global gyrokinetic calculations~\cite{P.Hill_1, Camenen_1}. 

Due to the radial variation of the equilibrium profiles, except for a critical value of flow shear, we expect the high growth rate isolated modes to be suppressed. Therefore, in general, the growth rate will be lower than the maximum value obtained from the local codes  even in the $n \rightarrow \infty$ limit. This might have important implications for the quasilinear predictions of the heat and particle fluxes. 

Finally, the procedure used in this work is quite general and can be used to explore more realistic tokamak equilibria. In our future work we will investigate the effect of shaping (eg elongation and triangularity) on the behaviour and stability of the linear global ITG modes. 
\section*{Acknowledgement}
The main author is extremely grateful to the Ministry of Higher Education in Kurdistan region of Iraq for the opportunity and funding they provided to study for a PhD at University of York. This work has also received funding from the European Union's Horizon 2020 research and innovation programme under grant agreement number 633053 and from the RCUK Energy Programme [grant number EP/I501045]. The views and opinions expressed herein do not necessarily reflect those of the European Commission. The simulations presented were carried out using supercomputing resources on HECToR (from the Plasma HEC Consortium EPSRC grant number EP/L000237/1).  A part of this work was also carried out using the HELIOS supercomputer system at Computational Simulation Centre of International Fusion Energy Research Centre (IFERC-CSC), Aomori, Japan, under the Broader Approach collaboration between Euratom and Japan, implemented by Fusion for Energy and JAEA. 


\end{document}